\documentclass{article}
\usepackage{spconf,amsmath,graphicx, multirow,url}

\title{QUARTZNET: DEEP AUTOMATIC SPEECH RECOGNITION WITH 1D TIME-CHANNEL SEPARABLE CONVOLUTIONS}

\name{\begin{tabular}{c}
Samuel Kriman$^{\dagger\star}$\qquad Stanislav Beliaev$^{\ddagger}$\sthanks{Work was conducted while S.Kriman and S.Beliaev were at NVIDIA}\qquad Boris Ginsburg\qquad Jocelyn Huang\\
Oleksii Kuchaiev\qquad Vitaly Lavrukhin\qquad Ryan Leary\qquad Jason Li\qquad Yang Zhang
\end{tabular}
}

\address{NVIDIA, USA \\
$^{\dagger}$Univ. of Illinois Urbana-Champaign, \ \ \ $^{\ddagger}$High School of Economics, Univ. of Saint Petersburg
}

\begin{document}
\maketitle

\begin{abstract}
We propose a new end-to-end neural acoustic model for automatic speech recognition. The model is composed of multiple blocks with residual connections between them. Each block consists of one or more modules with 1D time-channel separable convolutional layers, batch normalization, and ReLU layers. It is trained with CTC loss. The proposed network achieves near state-of-the-art accuracy on LibriSpeech and Wall Street Journal, while having fewer parameters than all competing models. We also demonstrate that this model can be effectively fine-tuned on new datasets. 
\end{abstract}%
\section{Introduction}
\label{sec:intro}

In the last few years, end-to-end (E2E) neural networks (NN) have achieved new state-of-the-art (SOTA) results on many automatic speech recognition (ASR) tasks. Such models replace the traditional multi-component ASR system with a single, end-to-end trained NN which directly predicts character sequences and therefore greatly simplify training, fine-tuning and inference. The latest E2E models also have very good accuracy, but this often comes at the cost of increasingly large models with high  computational and memory requirements.

The motivation of this work is to build an ASR model that achieves SOTA-level accuracy, while utilizing significantly fewer parameters and less compute power. Smaller models offer multiple advantages: (1) they are faster to train, (2) they are more feasible to deploy on hardware with limited compute and memory, and (3) they have higher inference throughput. 

We achieve this goal by building a very deep NN with 1D time-channel separable convolutions. This new network reaches near-SOTA word error rate (WER) on LibriSpeech \cite{panayotov2015librispeech} (see Table \ref{tab:LibriSpeech}) and WSJ \cite{wsj} (see Table \ref{tab:WSJ}) datasets with fewer than 20 million parameters, compared to previous end-to-end ASR designs which typically have  over 100 million parameters. We have released the source code and pre-trained models in the NeMo toolkit \cite{kuchaiev2019nemo}.\footnote{ \url{https://github.com/NVIDIA/NeMo}}

\section{Related Work}
\label{sec:relatedwork}
There has been a lot of work done in exploring compact network architectures and on investigating the trade-off between accuracy and size of neural networks, such as SqueezeNet \cite{iandola2016}, ShuffleNet \cite{zhang2018}, and EfficientNet \cite{tan2019}.
Our approach is directly related to MobileNets \cite{howard2017, sandler2018} and Xception \cite{Chollet_2017}, which uses depthwise separable convolutions \cite{Sifre2014RigidMotionSF, vanhoucke2019}. Each depthwise separable convolution module is made up of two parts: a depthwise convolutional layer and a pointwise convolutional layer. Depthwise convolutions apply a single filter per input channel (input depth). Pointwise convolutions are $1\times1$  convolutions, used to create a linear combination of the outputs of the depthwise layer. BatchNorm and ReLU are applied to the outputs of both layers.

Hannun et al \cite{hannun2019} applied a similar approach to ASR. They introduced an encoder-decoder model with time-depth separable (TDS) convolutions. The TDS model operates on data in time-frequency-channels ($T \times w \times c$) format, where $T$ is the number of time-steps, $w$ is the input width and $c$ is the number of channels.
The basic TDS block is composed of a 2D convolutional block with  $k \times 1$ convolutions over $(T \times w)$, and a fully-connected block, consisting of two $1\times1$ pointwise convolutions operating on $(w \cdot c)$ channels interleaved with layer-norm layers.  In contrast, in our work we operate on data in time-channel format ($T\times c$) and completely decouple the time and channel-wise parts of convolution. TDS block has  $k \times c^2 + 2\times (w \cdot c)^2$ parameters, while  QuartzNet model has $k \times c + c^2$ parameters, which allows for a dramatic reduction in model size while still achieving good WER.

Another very small ASR model  was introduced by Han et al \cite{han2019stateoftheart}, which uses multiple parallel streams of self-attention with dilated, factorized, although not separable, 1D convolutions. The parallel streams capture multiple resolutions of speech frames from the input by using different dilation rates per stream, and the results of the individual streams are concatenated into a final embedding. The best model has five streams with dilation rates 1-2-3-4-5. 
\section{Model architecture}
\label{sec:pagestyle}
\subsection{Basic model}
\begin{figure}[t]
 \centering
 \includegraphics[width=\linewidth]{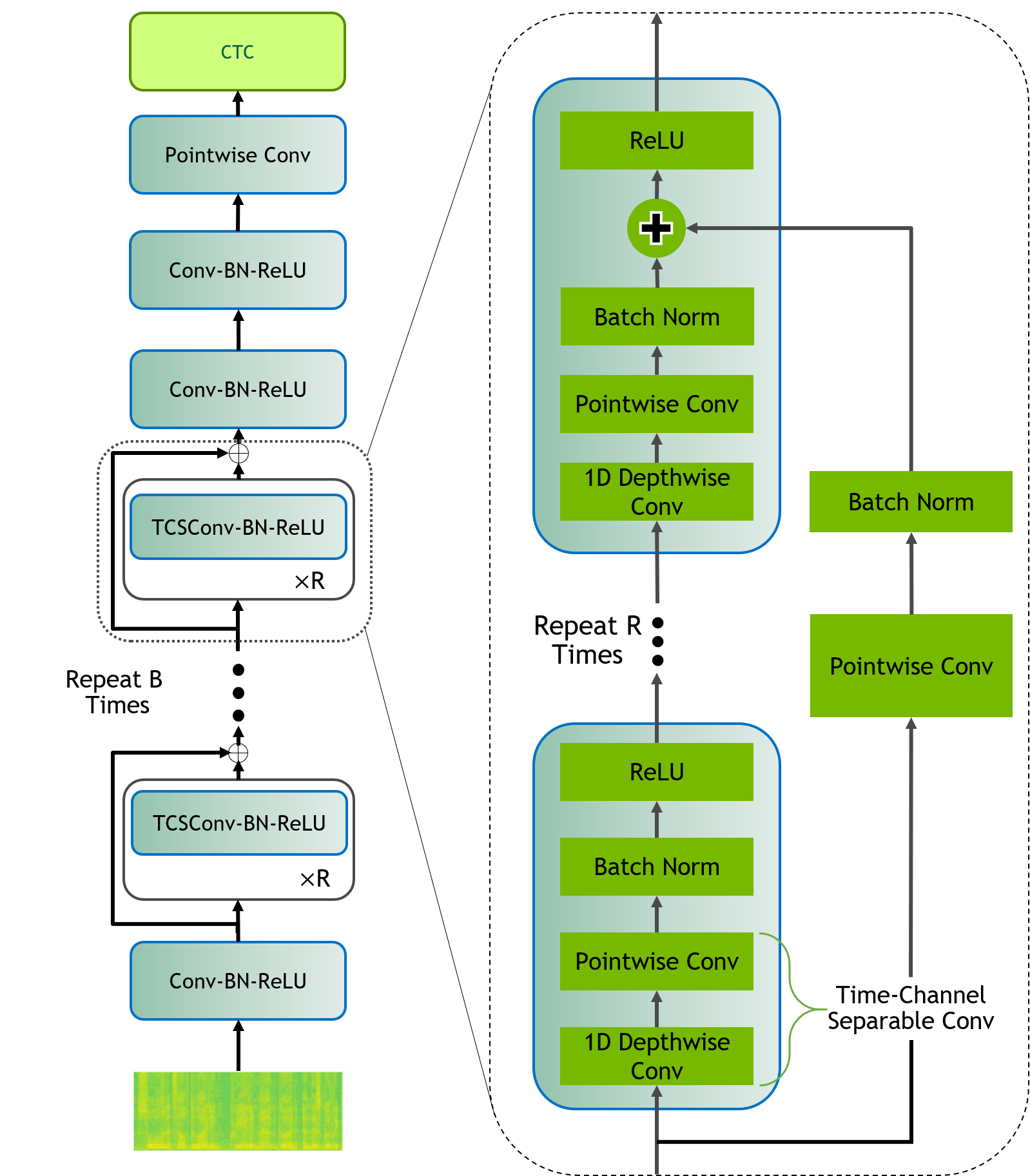}
 \caption{QuartzNet BxR architecture}
 \label{fig:quartz_arch}
\end{figure}

QuartzNet's design is based on the Jasper \cite{li2019jasper} architecture, which is a convolutional model trained with Connectionist Temporal Classification (CTC) loss \cite{graves2006}. The main novelty in QuartzNet's architecture is that we replaced the 1D convolutions with 1D time-channel separable convolutions, an implementation of depthwise separable convolutions. 1D time-channel separable convolutions can be separated into a 1D depthwise convolutional layer with kernel length $K$ that operates on each channel individually but \textbf{across $\pmb{K}$ time frames} and a pointwise convolutional layer that operates on each time frame independently but \textbf{across all channels}.

QuartzNet models have the following structure: they start with a 1D convolutional layer $C_1$ followed by a sequence of blocks. Each block $B_i$ is repeated $S_i$ times and has residual connections between blocks. Each block $B_i$ consists of the same base modules repeated $R_i$ times and contains four layers: 1) $K$-sized depthwise convolutional layer with $c_{out}$ channels, 2) a pointwise convolution, 3) a normalization layer, and 4) ReLU.  
The last part of the model consists of three additional convolutional layers ($C_2, C_3, C_4$). The $C_1$ layer has a stride of 2, and $C_4$ layer has a dilation of 2. 

Table \ref{tab:Model_Architectures} describes the QuartzNet-5x5, 10x5 and 15x5 models. There are five unique blocks across these models: $B_1$ - $B_5$. The different models repeat the blocks a different number of times, represented by $S_i$. QuartzNet-5x5 ($B_1-B_2-B_3-B_4-B_5$) has each group of blocks repeated 1 time, QuartzNet-10x5 ($B_1-B_1-B_2-B_2-...-B_5-B_5$) - repeated 2 times, and QuartzNet-15x5 ($B_1-B_1-B_1-...-B_5-B_5-B_5$) - repeated 3 times.

\begin{table}[t]
\caption{
QuartzNet Architecture. The model starts with a conv layer $C_1$ followed by a sequence of 5 groups of blocks. Blocks in the group are identical, each block $B_k$ consists of \textbf{R} time-channel separable \textbf{K}-sized convolutional modules with \textbf{C} output channels. Each block is repeated \textbf{S} times. The model has 3 additional conv layers ($C_2, C_3, C_4$) at the end.
}
\vspace{4pt}
\label{tab:Model_Architectures}
\centering
\scalebox{1.0}
{
\begin{tabular}{c c c c c c c c c}
 \hline
   \textbf{Block} & \textbf{R} & \textbf{K} & \textbf{C}& \multicolumn{3}{c}{\textbf{S}} \\
 & & & & \textbf{5x5} & \textbf{10x5} & \textbf{15x5} \\
 \hline
 $C_1$ & 1 & 33 & 256 & 1 & 1 & 1\\
 \hline
 $B_1$ & 5 & 33 & 256 & 1 & 2 & 3\\
 $B_2$ & 5 & 39 & 256 & 1 & 2 & 3\\
 $B_3$ & 5 & 51 & 512 & 1 & 2 & 3\\
 $B_4$ & 5 & 63 & 512 & 1 & 2 & 3\\
 $B_5$ & 5 & 75 & 512 & 1 & 2 & 3\\
 \hline
 $C_2$ & 1 & 87 & 512 & 1 & 1 & 1\\
 $C_3$ & 1 & 1 & 1024 & 1 & 1 & 1\\
 $C_4$ & 1 & 1 & $\|$labels$\| $   & 1 & 1 & 1\\
 \hline
 \textbf{Params, M}& & & & 6.7 & 12.8 & 18.9\\
 \hline
\end{tabular}
}
\end{table}

A regular 1D convolutional layer with kernel size $K$, $c_{in}$ input channels, and $c_{out}$ output channels has $K \times c_{in} \times c_{out}$ weights. The time-channel separable convolutions use $K \times c_{in} + c_{in} \times c_{out}$ weights split into $K \times c_{in}$ weights for the depthwise layer and $c_{in} \times c_{out}$ for the pointwise layer.

\begin{table}[th]
\centering
\caption{QuartzNet models with different depth trained on LibriSpeech for 300 epochs, greedy WER ($\%$).}
\vspace{4pt}
\label{tab:Depth}
\scalebox{1.0}
{
\begin{tabular}{c c c c} 
 \hline
 {\textbf{Model}} & {\textbf{dev-clean}} & {\textbf{dev-other}} & {\textbf{Params, M}} \\
 \hline
 5x5 & 5.39 & 15.69 & 6.7\\ 
 10x5 & 4.14 & 12.33 &  12.8 \\ 
 15x5  & 3.98 & 11.58 & 18.9\\ 
 \hline
\end{tabular}
}
\end{table}

The depthwise convolution is applied independently for each channel, so it contributes a relatively small portion of the total number of weights. This allows us to use much wider kernels, roughly 3 times larger than kernels used in wav2letter \cite{Wav2LetterV2} or Jasper \cite{li2019jasper} models.
We experimented with four types of normalization: batch normalization \cite{IoffeS15BatchNorm}, layer normalization \cite{Ba2016LayerNorm}, instance normalization \cite{ulyanov2016instance}, and group normalization \cite{Wu_2018_groupnorm}, and found that models with batch normalization have most stable training and give the best WER.

\subsection{Pointwise convolutions with groups}
The total number of weights for a time-channel separable convolution block is $K \times c_{in} + c_{in} \times c_{out}$ weights. Since $K$ is generally several times smaller than $c_{out}$, most weights are concentrated in the pointwise convolution part. In order to further reduce the number of parameters, we explore using group convolutions for this layer. We also added a group shuffle layer to increase cross-group interchange \cite{zhang2018}.

\begin{figure}[h]
 \centering
 \scalebox{1.0}{
 \includegraphics[width=\linewidth]{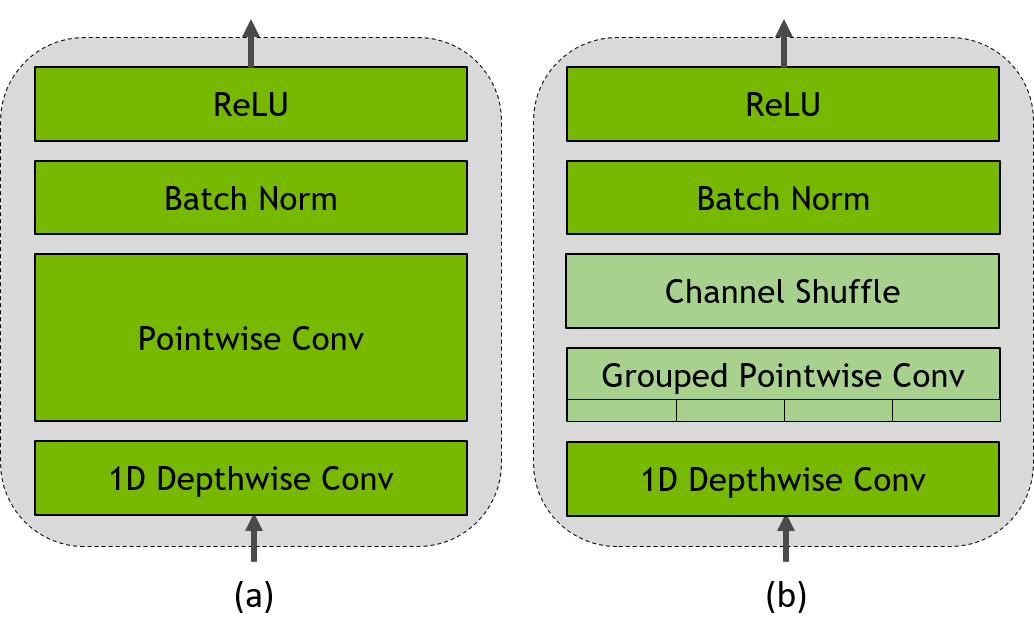}
 }
 \caption{(a) Time-channel separable 1D convolutional module (b) Time-channel separable 1D convolutional module with groups and shuffle}
 \label{fig:quartz_arch_groups}
\end{figure}

Using groups allows us to significantly reduce the number of weights at the cost of some accuracy. Table \ref{tab:GroupConv} shows the trade-off between accuracy and number of parameters for group sizes one, two, and four, evaluated on LibriSpeech. 

\begin{table}[thb]
\centering
\caption{QuartzNet-15x5 with grouped convolutions trained on LibriSpeech for 300 epochs, greedy WER ($\%$)}
\vspace{4pt}
\label{tab:GroupConv}
\begin{tabular}{c c c c} 
 \hline
 {\textbf{\# Groups}} & {\textbf{dev-clean}} & {\textbf{dev-other}} & {\textbf{Params, M}} \\
 \hline
 1  & 3.98 & 11.58 & 18.9 \\ 
 2  & 4.29 & 12.52 & 12.1 \\ 
 4  & 4.51 & 13.48 & 8.70 \\ 
 \hline
\end{tabular}
\end{table}

\section{Experiments}
We  evaluate QuartzNet's performance on LibriSpeech and WSJ datasets. We additionally experiment with  a transfer learning showcasing how a QuartzNet model trained with LibriSpeech and Common Voice \cite{CommonVoice} can be fine-tuned on a smaller amount of audio data, the WSJ dataset, to achieve better performance than training from scratch.

\subsection{LibriSpeech}
Our best results on the LibriSpeech dataset are achieved with the QuartzNet-15x5 model, consisting of 15 blocks with 5 convolutional modules per block (see Table \ref{tab:Model_Architectures}). By combining our network with independently trained language models (i.e., n-gram language models and Transformer-XL (T-XL) \cite{transformerxl}) we got WER comparable to the current SOTA.

The model with time-channel separable convolutions is much smaller than a model with regular convolutions and is less prone to over-fitting, so we use only data augmentation and weight decay for regularization during training. We experimented with  SpecAugment \cite{park2019},  SpecCutout, and speed perturbation \cite{Ko2015}. We achieved the best results with 10\% speed perturbation combined with Cutout \cite{cutout2017} which randomly cuts small rectangles out of the spectrogram.
The models are trained using NovoGrad optimizer \cite{novograd2019} with a cosine annealing learning rate policy. We also found that learning rate warmup  helps stabilize early training. 

\begin{table}[thb]
\centering
\caption{LibriSpeech results, WER ($\%$)}
\vspace{4pt}
\label{tab:LibriSpeech}
\scalebox{0.71}{
\begin{tabular}{cccccc} 
 \hline
 \multirow{2}{*}{\textbf{Model}}    &
 \multirow{2}{*}{\textbf{Augment}}  &
 \multirow{2}{*}{\textbf{LM}}       &
 \multicolumn{2}{c}{\textbf{Test}}  &
 \multirow{2}{*}{\textbf{Params,M}} \\
  &&& \textbf{clean}& \textbf{other}\\
 \hline
 wav2letter++ \cite{zeghidour2018} & speed perturb & ConvLM &  3.26 & 10.47 & 208 \\
  \hline
 LAS \cite{park2019}        & SpecAugment & RNN    & 2.5 & 5.8 & 360 \\
  \hline
 TDS Conv \cite{hannun2019} & dropout+    &  - & 5.36 & 15.64 & 37 \\
                            & label smooth & 4-gram & 4.21 &11.87 &  \\
                            &   & ConvLM & 3.28 & 9.84 &  \\
 \hline
 MSSA\cite{han2019stateoftheart} & speed perturb & 4-gram & 2.93 & 8.32 & 23\\
 &  & 4-LSTM & 2.20 & 5.82 & \\ 
 \hline
 JasperDR-10x5\cite{li2019jasper} & SpecAugment+  & -       & 4.32 & 11.82 & 333\\
                & speed perturb & 6-gram  & 3.24 & 8.76 & \\ 
                &               & T-XL    & 2.84 & 7.84 & \\ 
 \hline
 \hline
 QuartzNet 15x5 & SpecCutout+   & -      & 3.90 & 11.28 & 19 \\
                & speed perturb & 6-gram & 2.96 & 8.07 & \\
                &               & T-XL   & 2.69 & 7.25 & \\
 \hline
\end{tabular}
}
\end{table}
The training of the 15x5 model for 400 epochs
took $\approx 5$ days on one DGX1 server with 8 Tesla V100 GPUs with a batch size of 32 per GPU.  In order to  decrease the memory footprint as well as training time, we used mixed-precision training  \cite{micikevicius2017mixed}. We reduced the training time to just over four hours  by scaling training to SuperPod with 32 DGX2 nodes with larger number of epochs and with an increased global batch of 16K (see Table \ref{tab:librispeech_largebatch}).\footnote{Training even longer (3000 epochs) improved greedy WER on test-clean to 3.87\% and on test-other to  10.61\%.}

\begin{table}[!ht]
\centering
\caption{QuartzNet-15x5: large batch training on LibriSpeech, time to train (hours) and greedy WER ($\%$).}
\vspace{4pt}
\label{tab:librispeech_largebatch}
\centering
\scalebox{0.85}{
\begin{tabular}{cc cc cc cc} 
 \hline
 \multirow{2}{*}{\textbf{Batch}} &
 \multirow{2}{*}{\textbf{Epochs}} &
 \multirow{2}{*}{\textbf{Time, h}} &
 \multicolumn{2}{c}{\textbf{Dev}} &
 \multicolumn{2}{c}{\textbf{Test}} \\
  &&&\textbf{clean} & \textbf{other} & \textbf{clean} & \textbf{other} \\
  \hline
  256 & 400 & 122  & 3.83 & 11.08 & 3.90  & 11.28\\
  16K & 1500 & 4.3 &  3.71 & 10.78 & 4.04 &  11.06\\
  \hline
\end{tabular}
}
\end{table}

\subsection{Wall Street Journal}
We trained a smaller QuartzNet-5x3 model on the open vocabulary task of the Wall Street Journal dataset \cite{wsj}. 
We used train-si284 set for training, nov93-dev for validation, and nov92-eval for testing. 
The QuartzNet-5x3 model (see Table \ref{tab:WSJ_Architecture}) was trained for 1200 epochs with batch size 32 per GPU, data augmentation (10\% speed perturbation, SpecCutout)  and dropout of 0.2 using NovoGrad optimizer ($\beta_1=0.95$, $\beta_2=0.5$) with 1000 steps of learning rate warmup, a learning rate of 0.05, and weight decay 0.001. 

We used 2 external language models during inference: 4-gram (beam size=2048, alpha=3.5, beta=1.5) and Transformer-XL (T-XL). Both language models were constructed using only the official LM data of WSJ.

\begin{table}[thb]
\caption{QuartzNet-5x3 for WSJ. The model has the same layers $C_1, C_2, C_3, C_4$ as QuartzNet-15x5, but the middle part consists of only five blocks, each of which is repeated three times.}
\vspace{4pt}
\label{tab:WSJ_Architecture}
\centering
\scalebox{1.0}
{
\begin{tabular}{c c c c } 
 \hline
   \textbf{Block} & \textbf{R} & \textbf{K} & \textbf{C} \\
 \hline
 $C_1$ & 1 & 33 & 256 \\
 \hline
 $B_1$ & 3 & 63 & 512 \\
 $B_2$ & 3 & 63 & 512 \\
 $B_3$ & 3 & 75 & 512 \\
 $B_4$ & 3 & 75 & 512 \\
 $B_5$ & 3 & 75 & 512 \\
 \hline
 $C_2$ & 1 & 87 & 512 \\
 $C_3$ & 1 & 1 & 1024 \\
 $C_4$ & 1 & 1 & $\|$labels$\| $  \\
 \hline
\end{tabular}
}
\end{table}

We used following end-to-end models trained on standard speech features\footnote{Note, that wav2letter++ \cite{zeghidour2018b} with trainable front-end and convLM has even better WER: 6.8$\%$ for nov93-test, and 3.5$\%$ for nov92-dev. Here, we consider only models with a standard mel-filterbanks front-end.}
for comparison:\\
1) RNN-CTC \cite{Graves2014} -  CTC model with 5 bidirectional LSTM layers, 500 cells in each layer;\\
2) ResCNN-LAS \cite{Chorowski2016}: Listen-Attend-Spell model with deep residual convLSTM encoder and LSTM decoder + label smoothing;\\
3) Wav2Letter++ \cite{zeghidour2018b} - CTC model with 1D convolutional layers and instance norm.

\begin{table}[thb]
\centering
\caption{QuartzNet-5x3, WSJ, WER(\%)}
\vspace{4pt}
\label{tab:WSJ}
\scalebox{0.85}{
\begin{tabular}{c c c c c c} 
 \hline
 {\textbf{Model}} & {\textbf{LM}} & {\textbf{93-test}} & {\textbf{92-eval}} & {\textbf{ Params, M}} \\
 \hline
  RNN-CTC\cite{Graves2014}        & 3-gram & -   & 8.7 & 26.5\\
   \hline
  ResCNN-LAS\cite{Chorowski2016}  & 3-gram & 9.7 & 6.7 & 6.6\\
   \hline
  Wav2Letter++\cite{zeghidour2018b} & 4-gram  & 9.5 & 5.6 & 17\\
   & convLM  & 7.5 & 4.1 &  \\
 \hline
 \hline
  QuartzNet-5x3 & 4-gram & 8.1  & 5.8 & 6.4 \\
   & T-XL & 7.0 & 4.5 &  \\
 \hline
\end{tabular}
}
\end{table}

\subsection{Transfer Learning}
As our model is smaller than other models, we were interested in how well it could learn to generalize to data from various sources, especially if the amount of target speech is much smaller than the training data. Our setup consists of training QuartzNet 15x5 on a combination of LibriSpeech \cite{panayotov2015librispeech} and Mozilla's Common Voice \cite{CommonVoice}\footnote{We used the validated set of Common Voice, ver. en\_1087h\_2019-06-12.} datasets, and then fine-tuning this trained model on the 80 hour WSJ dataset.  Table \ref{tab:TransferLearning} shows the WER achieved on LibriSpeech prior to fine-tuning and the result on WSJ after fine-tuning.

\begin{table}[!ht]
\centering
\caption{QuartzNet15x5 transfer learning. The model was pre-trained of LibriSpeech-train  and Mozilla’s Common Voice datasets,  and fine-tuned on the 80-hour WSJ dataset. The model was evaluated on LibriSpeech and WSJ, WER($\%$).}
\vspace{4pt}
\label{tab:TransferLearning}
\scalebox{0.85}{
\begin{tabular}{c c c c c} 
 \hline
{\multirow{2}{*}{\textbf{LM}}} &
\multicolumn{2}{c}{\textbf{LibriSpeech}} &
\multicolumn{2}{c}{\textbf{WSJ}}\\
& \textbf{test-clean}  & \textbf{test-other} & \textbf{93-test} & \textbf{92-eval}\\
 \hline
- & 4.19 & 10.98 & 8.97 & 6.37\\ 
4-gram & 3.21 & 8.04 & 5.57 & 3.51\\ 
T-XL & 2.96 & 7.53 & 4.82 & 2.99 \\ 
 \hline
\end{tabular}
}
\end{table}

\section{Conclusions and future directions}
\label{sec:conclusions}
We introduced a new end-to-end speech recognition model, based on deep neural network with 1D time-channel separable convolutional layers. The model showed close to state-of-the art performance on Wall Street Journal and on LibriSpeech while being significantly smaller than all other end-to-end systems with similar accuracy. The small model footprint opens new possibility for speech recognition on mobile and embedded devices.

This work described a CTC-based model, but we are exploring models where the QuartzNet encoder is combined with attention-based decoders.


\bibliographystyle{IEEEbib}
\bibliography{refs}

\end{document}